\documentclass{elsart5p}
\usepackage{graphicx,natbib,amssymb}
\journal{New Astronomy}
\newcommand{\kpc}{\,{\rm kpc}}
\newcommand{\kms}{\,km\,s$^{-1}$}
\newcommand{\cmt}{\,cm$^{-3}$}   
\newcommand{\cmd}{\,cm$^{-2}$}   
\newcommand{\myr}{\,$M_{\odot}\,{\rm yr}^{-1}$}
\newcommand{\es}{$\,\rm erg\,s^{-1}$}
\newcommand{\ecs}{$\,\rm erg\,cm^{-2}\,s^{-1}$}
\newcommand{\ecsa}{$\,\rm erg\,cm^{-2}\,s^{-1}\,\AA^{-1}$}
\newcommand{\ha}{H$\alpha$}
 
\newcommand{\hi}{H\,{\small I}}

\newcommand{\ro}{\,$R_{\odot}$}
\newcommand{\mo}{\,$M_{\odot}$}
\newcommand{\lo}{\,$L_{\odot}$}
\def\I{\rm {\scriptsize I}}

\begin{document}
\begin{frontmatter}

\title{Multiwavelength modelling the SED of supersoft X-ray sources \\[2mm]
       II. RS~Ophiuchi: From the explosion to the SSS phase}

\author{A.~Skopal\thanksref{fn1}, \thanksref{fn2}}

\thanks[fn1]{E-mail: skopal@ta3.sk}
\thanks[fn2]{Visiting Astronomer: Astronomical Institut, Bamberg}
\thanks[fn3]{doi:?}

\address{Astronomical Institute, Slovak Academy of Sciences,
         059\,60 Tatransk\'{a} Lomnica, Slovakia \\[1mm]
        {\rm Received 16 July 2013; accepted 14 December 2013}
        }
\vspace*{5mm}

{\small
\hspace*{-15cm} 
{\rm H I G H L I G H T S}\\
\begin{itemize}
\item
Physical parameters of the nova RS Oph were determined from its X-ray/IR spectrum. 
\item
The model revealed the presence of a strong stellar and nebular component. 
\item
The luminosity was super-Eddington during the whole burning phase. 
\item
The high luminosity was sustained by a super-critical accretion. 
\end{itemize}
}
\vspace*{-7mm}

\begin{abstract}
RS~Oph is a recurrent symbiotic nova that undergoes nova-like 
outbursts on a time scale of 20 years. Its two last eruptions 
(1985 and 2006) were subject of intensive multiwavelengths 
observational campaign from the X-rays to the radio. 
This contribution aims to determine physical parameters and 
the ionization structure of the nova from its explosion to 
the first emergence of the supersoft X-rays (day 26) by using 
the method of multiwavelength modelling the SED. 
From the very beginning of the eruption, the model SED revealed 
the presence of both a strong stellar and nebular component 
of radiation in the spectrum. 
During the first 4 days, the nova evinced a biconical ionization 
structure. The $\sim 8200$\,K warm and 160--200\ro\ extended 
pseudophotosphere encompassed the white dwarf (WD) around its 
equator to the latitude $>40^{\circ}$. The remaining space around 
the WD's poles was ionized, producing a strong nebular continuum 
with the emission measure $\textsl{EM} \sim 2.3 \times 10^{62}$\cmt\ 
via the fast wind from the WD. 
The luminosity of the burning WD was highly super-Eddington for 
the whole investigated period. 
The wind mass loss at rates of $10^{-4} - 10^{-5}$\myr\ and 
the presence of jets suggest an accretion throughout a disk 
at a high rate, which can help to sustain the super-Eddington 
luminosity of the accretor for a long time. 
%
\end{abstract}
\end{frontmatter}
%
%
\section{Introduction}

The star RS~Ophiuchi (RS~Oph) is located in the direction 
nearby to the galactic center 
($l_{\rm II} = +19^{\circ}.8 ~ b_{\rm II} = +10^{\circ}.4$) 
at a distance $d = 1.6 \pm 0.3\kpc$ 
\citep[][]{hjell+86,bode87,barry08a}. 
The light of RS~Oph is significantly attenuated by 
interstellar extinction. \cite{hjell+86} determined from 
\hi\ 21\,cm measurements the interstellar absorbing 
column, $N_{\rm H} = (2.4\pm 0.3)\times 10^{21}$\,\cmd, 
and \cite{snijders87} derived the color excess 
$E_{\rm B-V} = 0.73\pm 0.1$\,mag from ultraviolet 
\textsl{IUE} spectra. 
According to the relationship between $N_{\rm H}$ and 
$E_{\rm B-V}$ \citep[e.g.][]{d+s94}, both these parameters 
are consistent.
%

RS~Oph is a recurrent symbiotic nova comprising a late-type, 
K7\,III, giant \citep{ms99} and a WD with a mass close to the 
Chandrasekhar limit \citep[e.g.][and references therein]{bode87}, 
in a 456-day orbit \citep{f+00,brandi+09}. 
Its nova-like outbursts are characterized with brightening
by about 7\,mag and a recurrence period of about 20 years.
Historically, 6 eruptions have been recorded unambiguously.
The first one in 1898 and the last one on 2006 February 12.83
\citep[][]{evans88,narumi+06}. 
The recurrence period of approximately 20 years and a bright
peak magnitude, $V = 4-5$, made RS~Oph a good target for
multifrequency observational campaigns 
\citep[e.g.][and references therein]{evans07}. 

At the beginning of the last two outbursts of 1985 and 2006, 
observations revealed a non-spherical shape of the nova 
ejecta, 
(i) on the radio maps 
    \citep[e.g.][]{taylor+89,tob+06,rupen+08,sok+08}, 
(ii) by interferometric technique in the near and mid-IR 
    \citep[e.g.][]{monnier+06,lane+07,chesneau+07,barry08b}, 
(iii) by the \textsl{HST} imaging in the optical \citep{bode+07}, 
(iv) by the \textsl{Chandra} X-ray Observatory detected 
     as an extended X-ray emission elongated in the line with 
     the extended infrared emission \citep[][]{luna+09} and 
(v) by the asymmetric line profiles measured on day 13.9, 
    interpreted by \cite{drake+09} as a result of the shock 
    collimation due to equatorial circumstellar density 
    enhancement. 
Particularly, highly collimated jet-like ejection was 
indicated by the optical spectroscopy during the first 30 
days after the eruption \citep[][]{sk+08a,banerjee}. 

The 2006 outburst was intensively monitored with the 
{\em Swift} and the {\em Rossi X-ray Timing Explorer}. 
Prior to day 26, the behaviour of the X-ray flux was 
described by the evolution of shock systems established as 
the high-velocity ejecta impacted the red giant wind 
\citep[][]{bode+06,sok+06}. The evolving shock in the 2006 
outburst of RS~Oph was analyzed in detail by \cite{ness+09} 
using the high-resolution X-ray grating observations with 
\textsl{Chandra} and \textsl{XMM-Newton}. The X-ray 
observations demonstrated that the supersoft X-ray phase of 
nova RS~Oph started rapidly from day 26 
\citep[see also][]{osborne+11}. 

The detailed spectral evolution of the 2006 outburst in the optical 
was reported by \cite{iijima09}. He suggested that the profiles 
of the prominent emission lines resulted from a rapidly expanding 
($\sim 1000$\kms), low density part and a slowly moving 
($\sim 100$\kms) high density part in the ejecta, which are 
probably related to the expansion along the polar and the 
equatorial regions, respectively. 
%

In this contribution I aim to determine physical parameters 
of the nova RS~Oph from its explosion to the first emergence 
of the supersoft X-rays (day 26) by using the method of 
multiwavelength modelling the SED. Based on the results from 
spectral fits, I also explore the ionization structure of the 
nova during the first 4 days of its outburst. 
Section~2 introduces multi-band observations at selected days, 
and the results of their SED-fitting analysis are given in 
Sect.~3. Discussion and summary are found in Sects.~4 and 5, 
respectively. 

\section{Observations}

To map the evolution of the fast nova RS~Oph by the multiwavelength 
approach, simultaneous observations covering a large wavelength 
range, taken at different days after the explosion, are required. 
Rapid changes in the SED following the blast require frequent 
sets of observations. According to these conditions, it was 
possible to select (near-)simultaneous observations at/around 
day 1, 3.8, 7.3, 14.6, 19.5 and 26 after the optical maximum of 
the RS~Oph outburst. 
Selection of the appropriate data benefited from a strong 
similarity of the recent 1985 and 2006 outbursts in the 
supersoft X-ray, optical and near-IR light curve profiles 
\citep[e.g.][]{ness+07,ros87,banerjee}. Therefore, observations 
taken at the same time after the optical maximum of these 
outbursts were assumed to be simultaneous. 

The UV to near-IR observations were dereddened with 
$E_{\rm B-V}$ = 0.73 and the resulting parameters were scaled 
to the distance of 1.6\kpc\ (see Sect.~1). 
The log of the used observations is given in Table~1. 
%
%
\begin{table}
\caption[]{Log of observations}
\begin{center}
\begin{tabular}{cccc}
\hline
\hline
  Julian date  &~~Day$^{a}$  & Region  & Observatory~/~(ref.) \\
\hline
2\,453779.97 &~~0.64  &$J,H$     & private~/~(1)    \\
2\,453779.99 &~~0.66- &$B,V,R_{\rm C},I_{\rm C}$ & private~/~(2) \\
2\,453780.71 &~~-1.38 &$-$       & private~/~(2)      \\
2\,453780.97 &~~1.64  &$J,H$     & private~/~(1)   \\
2\,453780.33 &~~1     &$J,H,K,L$&\textsl{SAAO$^b$,Mt.~Abu}$^b$/~(3) \\
2\,453780.40 &~~1.07  & 350--920\,nm& \textsl{HCT, VBT}/~(4)\\
\hline
2\,453783.13 &~~3.8 & 8.06--12.26\,$\mu$m &\textsl{Keck N-band}/~(5)\\
2\,453783.33 &~~3-4 & 350--930\,nm        &\textsl{HCT, VBT}/~(4)\\
2\,453783.13 &~~3.8 & $B,V,R_{\rm C},I_{\rm C}$&\textsl{VSOLJ}$^c$\\
2\,453783.13 &~~3.8 & $J,H,K,L$ &\textsl{SAAO$^c$,Mt.~Abu}$^c$/~(3)\\
\hline
2\,446099.30 &~~7.33 & 115--335\,nm  &\textsl{IUE} \\
2\,446099.30 &~~7.33 & $UVI_CHKL$    &\textsl{SAAO}$^c$   \\
\hline
2\,446106.61 &~~14.6 & 115--335\,nm  &\textsl{IUE} \\
2\,446106.61 &~~14.6 & $UVI_CHKL$    &\textsl{SAAO$^c$,Mt.~Abu}$^c$/~(3)\\
\hline
2\,446111.51 &~~19.5 & 115--335\,nm  &\textsl{IUE} \\
2\,446111.51 &~~19.5 & $UVI_CHKL$    &\textsl{SAAO$^c$,Mt.~Abu}$^c$/~(3)\\
\hline
2\,453805.53 &~~26.20 &  0.65--3.1\,nm  &\textsl{XMM-Newton}  \\
2\,446117.96 &~~25.99 &  115--335\,nm  &\textsl{IUE}        \\
2\,446117.96 &~~25.99 & $UVI_CJHKL$    &\textsl{SAAO}$^c$   \\
\hline
\hline
\end{tabular}
\end{center}
$^{a}$~=\,JD -- JD$_{\rm max}$ 
         (JD$_{\rm max}$~2\,446\,091.97 as on 1985 Jan. 26.47; \\
\hspace*{23mm}
          JD$_{\rm max}$~2\,453\,779.33 as on 2006 Feb. 12.83) \\
$^{b}$~values extrapolated to day 1 \\
$^{c}$~values interpolated to the given day\\
1 - \cite{west}, 
2 - \cite{sostero} and VSOLJ observers (see Fig.~1), 
3 - \cite{evans88,banerjee}, 
4 - \cite{anupama08}, 
5 - \cite{barry08b}. 
\end{table}

\subsection{Observed SED of the giant}

Radiation of the giant in RS~Oph can be measured directly 
only during the quiescent phases. The preliminary model SED 
showed that the giant dominates the spectrum from the V passband 
to longer wavelength during the post-outburst minimum 
\citep[see Fig.~1 of][ day 253]{sk+08a}. 
Therefore, the giant's continuum was approximated by the 
photometric $VR_CI_C$ fluxes from the 2006 post-outburst 
minimum (the data were collected by the VSOLJ observers, 
Kiyota, Kubotera, Maehara and Nakajima), $JHKL$ flux-points 
obtained during 1986 quiescence \citep{evans88} and the 
archival $ISO$ spectrum from 23/09/1996 which determines the 
region between 2.3 and 4.0\,$\mu$m. Some representative fluxes 
were taken from the infrared (0.8--5\,$\mu$m) spectra 
obtained during 2006 August to 2008 July 
\citep[see Fig.~6 of][]{rushton+10}. 
%
%
\begin{figure*}[!t]
\centering
\begin{center}
%
\resizebox{13cm}{!}{\includegraphics[angle=-90]{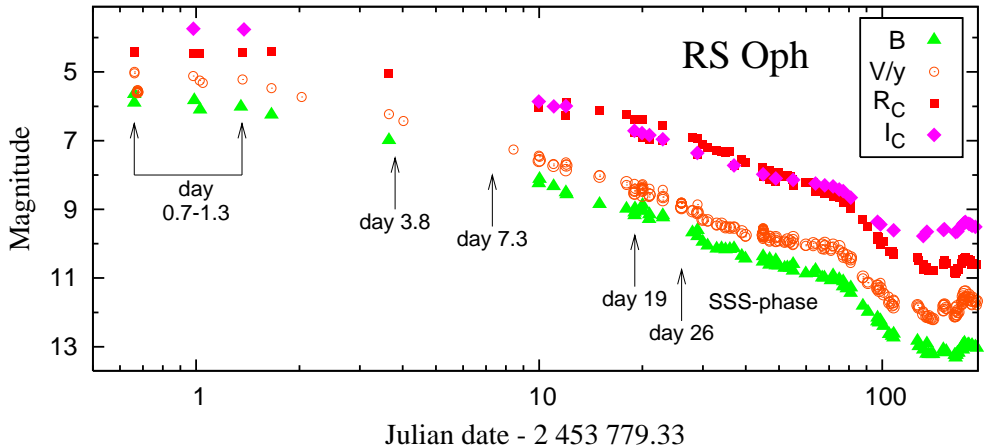}}
\end{center}
\caption[]{
The $BVR_CI_C$ light curves of RS~Oph from its optical maximum 
(2006 February 12.83) to the quiescent phase (d$>100$). 
Arrows denote times with reconstructed SEDs (Table~1, 
Fig.~\ref{fig:seds}). The data were collected by the VSOLJ 
observers, Kiyota, Kubotera, Maehara and Nakajima. 
          }
\label{fig:lcs}
\end{figure*}

\subsection{Day 0.7--1.4}

Observations at the very beginning stage of the nova outburst 
were restricted to $B, V, R_C, I_C$ flux-points measured 
between day 0.66 and 1.37
 \citep[][]{sostero} and $J$ and $H$ 
magnitudes obtained by \cite{west} at day 0.64 and 1.64 after 
the optical maximum of the 2006 outburst. In addition, fluxes 
derived from $K$ and $L$ magnitudes, obtained by linear 
extrapolation of the data measured at day 8.68, 9.65, 15.65 by 
\cite{evans88} and at day 11.12, 15.09, 17.12 by \cite{banerjee} 
to day 1, were used. 

\subsection{Day 3.8}

Approximately 3.8 days following the 2006 outburst, RS~Oph was 
observed using the Keck Interferometer Nuller (KIN), which 
operates in $N$ band from 8 to 12.5\,$\mu$m \citep[][]{barry08b}. 
Corresponding fluxes, used in this paper, represent the light 
from the entire region detected by the KIN in its inner and 
outer spatial regime (Barry, private communication). 
At/around day 4, on 2006 February 16, a low resolution spectrum 
($R \sim 1000$) covering the range between 350 and 920\,nm was 
obtained at the Indian Astronomical Observatory with the 2-m 
Himalayan Chandra Telescope \citep[][]{anupama08}. The spectrum 
is available 
at\footnote{http://www.astro.keele.ac.uk/rsoph/pdfs/index.htm}. 
For the purpose of this paper we used its short wavelength part 
covering the Balmer jump. We calibrated the spectrum to absolute 
fluxes with the aid of the simultaneous photometric observations. 
Photometric $B, V, R_C, I_C$ magnitudes were collected by 
the VSOLJ observers. Flux in the $I_C$ band was estimated by 
the interpolation to day 3.8 (see Fig.~\ref{fig:lcs}). 
The near-IR $J$ and $H$ photometric fluxes at day 3.8 were 
obtained by linear interpolation of the values published 
by \cite{west}, \cite{evans88} and \cite{banerjee}. 
The $K$ and $L$ magnitudes were obtained by the same way as 
described in Sect.~2.2, but extrapolating the measured 
values to day 3.8. 

\subsection{Days 7.3--19.5}

The first ultraviolet spectra od RS~Oph were taken at day 7.3
by the \textsl{IUE} satellite (1985 February 2). We used the 
spectra SWP25152 and LWP05277, exposed in the low dispersion 
mode using the large aperture. The LWP05277s spectrum, which 
was taken with the small aperture, was used to fill in the 
saturated part of the large aperture spectrum. The small 
aperture spectrum was multiplied by a factor of 2.2 to match 
the large aperture one. 
The following \textsl{IUE} spectra exposed in both primes were 
made on 1985 February 10, at day 14.6 (SWP25209, LWP5335-6). 
%
%
The last set of the well exposed \textsl{IUE} spectra, without 
measured SSS X-ray counterpart, 
was obtained at day 19.55 (SWP25246, LWP05367). The spectra 
were complemented with the $UVI_CHKL$ fluxes, derived from 
the photometric measurements 
\citep[][and Fig.~1]{evans88,banerjee}. 

\subsection{Day 26}

The first detection of supersoft X-ray emission was made by 
the \textsl{X-Ray Telescope} (0.3--10\,keV) onboard the 
{\em Swift} satellite on day 26.0 
\citep[see Fig.~3 of][]{osborne+11} and by the 
\textsl{XMM-Newton} satellite, 26.1 days after the 2006 
eruption \citep[see Fig.~1 of][]{ness+09}. 
Simultaneously, at day 26.0 after the optical maximum of the 
1985 outburst, the ultraviolet spectra (SWP25289, LWP05403) 
were obtained by the \textsl{IUE} satellite. 
Spectroscopic observations were complemented with the 
$UVI_CJHKL$ photometric fluxes as in previous cases. 
Fluxes derived from $B$ and $R_C$ magnitudes were not used in 
the modelling the continuum SED, because of a strong saturation 
with emission lines \citep[see][]{sk07}. 

The X-ray-grating spectrum (RGS1) was extracted using the standard 
tools provided by the mission-specific software packages 
(Hanke, private communication). The spectrum was described 
by \cite{nelson+08} in detail. 
To correct the observed X-ray fluxes for absorptions, the 
{\em tbabs} absorption model for ISM composition with abundances 
given by \cite{wilms+00} (e.g. $\log(A_{\rm OI})+12 = 8.69$) 
was used. 
%
%
\begin{figure*}[!t]
\centering
\begin{center}
\resizebox{12cm}{!}{\includegraphics[angle=-90]{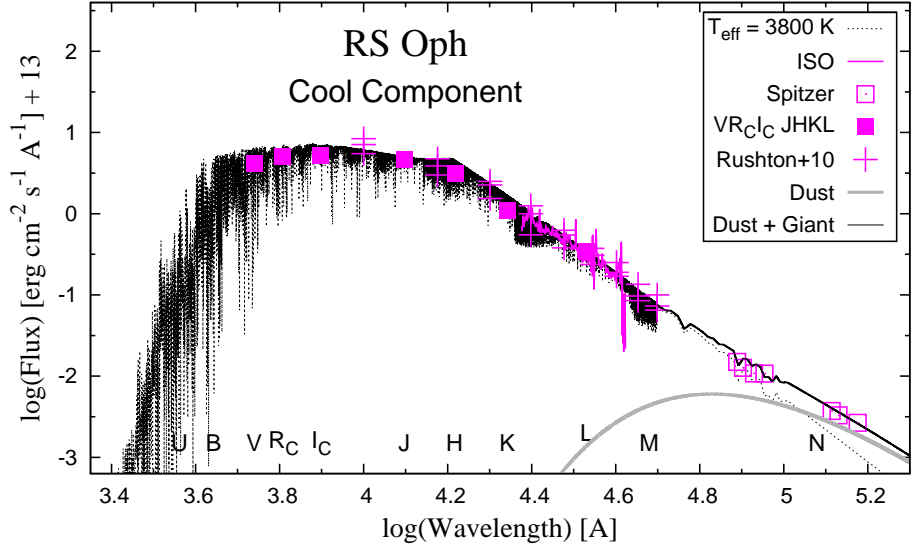}}
\end{center}
\caption[]{
The observed SED of the giant in RS~Oph (in violet) compared 
with an atmospheric model calculated for $T_{\rm eff} = 3800$\,K 
(dotted line). A weak blackbody radiating at $\sim 450$\,K is 
required to fit the continuum around the $N$ band 
\citep[data from][]{barry08b}. 
          }
\label{fig:giant}
\end{figure*}

\section{Results of the SED-fitting analysis}

In this section I describe modelling the observed SEDs at the 
selected days after the optical maximum (Sect.~2). 
The fitting analysis was performed in the same way as introduced 
by Skopal (2014, Paper~I). The best solution was selected from 
a grid of model parameters, which corresponded to a minimum of 
a standard $\chi^2$ function. Also in this paper, typical 
errors in the measured fluxes of $\sim10$\%  were adopted. 
Because of the presence of the giant in the binary, the spectrum 
$F(\lambda)$ emitted by RS~Oph was disentangled into three 
components, 
%
%
%
\begin{equation}
 F(\lambda) = F_{\rm h}(\lambda) +
              F_{\rm n}(\lambda) +
              F_{\rm g}(\lambda), 
\label{eq:fmod}
\end{equation}
%
%
where $F_{\rm h}(\lambda)$ is the flux produced by the WD 
pseudophotosphere, $F_{\rm n}(\lambda)$ is the nebular 
component from thermal plasma and $F_{\rm g}(\lambda)$ represents 
the contribution from the giant. They are defined in Paper~I. 
To simplify the task, we first estimated the contribution from 
the giant using observations made during quiescent phase 
(Sect.~2.1). Then, assuming that the giant's radiation is 
constant, its contribution was subtracted from the observed 
fluxes. 
Resulting model SEDs are depicted in Figs.~\ref{fig:giant} and 
\ref{fig:seds}, and the corresponding  parameters are collected 
in Table~2. 

\subsection{Radiation characteristics and radius of the giant}

Empirical relations between the spectral type (K7\,III, Sect.~1)
and effective temperature ($T_{\rm eff}$) for cool giants suggest
$T_{\rm eff} \sim 4000$\,K and a linear radius 
$R_{\rm g} \sim 50$\ro\ for the giant in RS~Oph
\citep[e.g.][]{bel+99}. 

According to the surface brightness relation for M-giants
\citep[][]{ds98}, the reddening-free magnitudes from the 1986 
quiescence, \citep[$J$ = 7.09 and $K$ = 6.41\,mag,][]{evans88}
yield the angular radius of the giant 
$\theta_{\rm g} = R_{\rm g}/d = 7.0\times 10^{-10}$, i.e.
$R_{\rm g} = 50\,(d/1.6\kpc)$\ro. 

Modelling the infrared spectra of RS~Oph, taken during its 
quiescent phase (2006 August -- 2008 July), \cite{pavlenko+08} 
and \cite{rushton+10} determined the effective temperature 
of the giant $T_{\rm eff} = 4100 \pm 100$\,K and 
$4200 \pm 200$\,K, respectively. 

Using our approach (see Sect.~2.5 in Paper~I), the observed 
SED of the giant (550--5000\,nm, Sect.~2.1) can be compared 
with the synthetic spectra \citep[models from][Fig.~2 here]{h+99} 
with limiting $T_{\rm eff}$ = 3800 and 4000\,K, which are scaled 
with the angular radius of the giant, 
$\theta_{\rm g} = R_{\rm g}/d$ = 8.6 and $7.7\times\,10^{-10}$, 
respectively (see Eq.~(10) in Paper~I). These couples of 
parameters define the observed bolometric flux 
$F_{\rm g}^{\rm obs} = \theta_{\rm g}^2 \sigma T_{\rm eff}^4 
= (8.7 \pm 1.2)\times 10^{-9}$\,\ecs, where the uncertainty 
represents that in the observed fluxes ($\sim 15$\%). 
The distance $d$ converts this quantity to the luminosity 
of the giant $L_{\rm g} = (690 \pm 95)\,(d/1.6\kpc)^2$\lo, 
and $\theta_{\rm g}$ to the radius of the giant 
$R_{\rm g} = (61 - 55)\,(d/1.6\,\kpc)$\ro. 
Figure~\ref{fig:giant} compares the model for 
$T_{\rm eff}$ = 3800\,K and 
$\theta_{\rm g} = 8.6\times\,10^{-10}$. 

The giant underfills its Roche lobe for the distance of 1.6\kpc. 
The orbital solution for both components of the binary 
\citep[][]{brandi+09} and the WD mass of 1.3\mo\ yield 
the Roche lobe radius of the giant $\sim 110$\ro, which is 
a factor of $ \sim 2$ larger than $R_{\rm g}$ (see also 
Sect~4.7 in the following paper, {\em RS~Ophiuchi: 
The supersoft X-ray phase and beyond}). 

\subsection{Optical maximum: The first day}

The reconstructed SED from around day 1 covers the range of 
0.44--3.4\,$\mu$m (Sect.~2.2). Fitting these data revealed 
the presence of two strong radiative contributions within 
this domain. A stellar-type component from the inflated WD's 
pseudophotosphere and that from thermal nebula. The former 
dominated the SED within the $B, V, R_C, I_C, J, H$ passbands, 
while the latter was larger within the $K$ and $L$ bands 
(see Fig.~\ref{fig:seds}). 
As a result, the best model SEDs corresponded to a relatively 
small range of the photospheric temperatures, 
$T_{\rm h} = 10000 - 9400$\,K, but to a large range of electron 
temperatures, $T_{\rm e} = 15000 - 40000$\,K. 
This is because of a weak dependence of the slope of the nebular 
continuum on $T_{\rm e}$ for $\lambda \gtrsim 1\,\mu$m 
and the too short part of the observed SED with dominant nebular 
radiation. The fits correspond to large values of the reduced    
$\chi^2$ sum ($\chi^2_{\rm red} = 2.93 - 2.86$), which is given 
mainly by the uncertainty in the extrapolated values of $K$ and 
$L$ flux-points. 
As no considerable changes in $T_{\rm e}$ of the extended nebula 
can be expected within a few days only, we fixed its value 
to 22000\,K, obtained from modelling the 3.8-day SED, which 
is well defined (see Sect.~3.3). This assumption constrains 
$T_{\rm h} = 9,600 \pm 300$\,K and the angular radius of the 
hot stellar source $\theta_{\rm h} = (2.2\pm 0.2)\times 10^{-9}$ 
($\chi^2_{\rm red} = 2.9$). These model parameters yield the 
effective radius of the WD's pseudophotosphere, 
  $R_{\rm h}^{\rm eff} = (156 \pm 14)(d/1.6\kpc)$\ro\ 
and its bolometric luminosity 
  $L_{\rm h} = (7.2 \pm 2)\times 10^{38}(d/1.6\kpc)^2$\es. 
The scaling factor of the nebular component of radiation 
$k_{\rm n} = 1.0\times 10^{18}$\,cm$^{-5}$ 
corresponds to a huge emission measure 
$EM = 3.2\times 10^{62}(d/1.6\kpc)^2$\cmt\ 
(see Eqs.~(8) and (9) in Paper~I). 

Finally, I note that the presence of a strong nebular component 
in the spectrum at the very beginning of the outburst was 
also supported by spectroscopic observations of \cite{banerjee}, 
who measured broad hydrogen emission lines of the Paschen and 
Bracked series with indication of the Bracket jump in emission 
at day 1.16. 

\subsection{Optical maximum: Day 3.8}

\begin{figure*}[!t]
\centering
\begin{center}
%
\resizebox{15.75cm}{!}{\includegraphics[angle=-90]{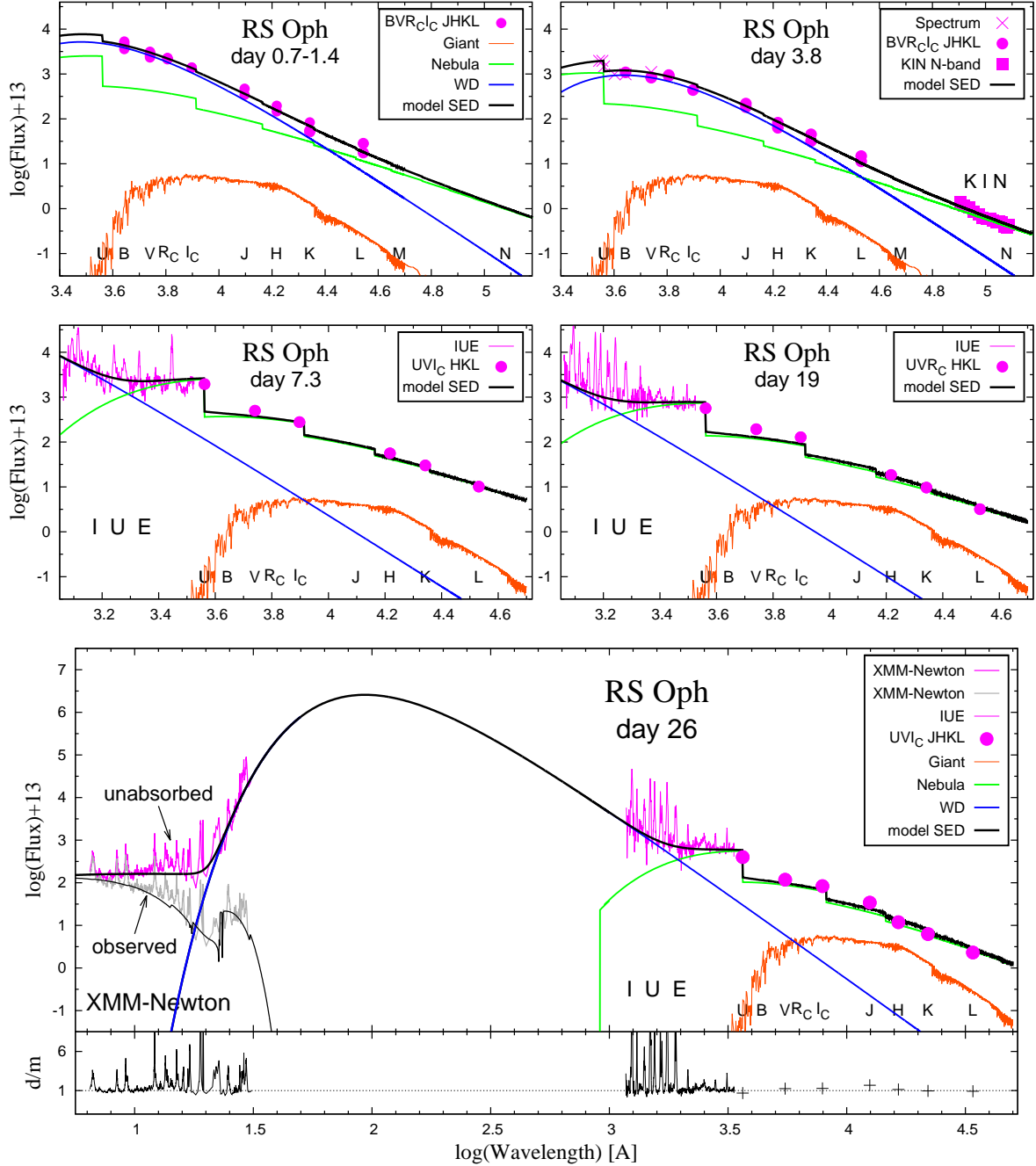}}
\end{center}
\caption[]{
A comparison of the measured SED (in violet) and model SED (black 
solid line) of RS~Oph from day 1 after its explosion to day 26. 
Data-to-model ratios (d/m) are plotted for the day-26 SED 
to judge the reliability of the fit. 
Models are described in Sect.~3 and their parameters given 
in Table~2. Fluxes are in units \ecsa. 
          }
\label{fig:seds}
\end{figure*}

Both the Keck N-band and the optical spectroscopic observations 
confirmed independently the presence of a strong nebula in the 
spectrum. The former by the slope of the $N$-band fluxes, and 
the latter by the Balmer jump in emission. On the other hand, 
photometric $B, V, R_C, I_C, J, H, K$ flux-points were dominated 
by the stellar component of radiation (Fig.~\ref{fig:seds}). 
These signatures and sufficiently large wavelength interval of 
the measured fluxes ($\sim$0.35--12.26\,$\mu$m), allowed to 
fit observations unambiguously. The stellar component of 
radiation became even cooler ($T_{\rm h} = 6700 \pm 200$\,K) 
than during day 1, and the effective radius of its source 
enlarged to $163 \pm 14 (d/1.6\kpc)$\ro\ that corresponds to 
$L_{\rm h} = (2 \pm 0.5)\times 10^{38}(d/1.6\kpc)^2$\es. 
The nebular emission from hydrogen and helium declined by 
a factor of $\sim$2.5 with respect to day 1, but still was 
very strong, keeping its $EM$ as high as 
$1.3\times 10^{62}$\,\cmt. Its bolometric luminosity 
for the Case~B (i.e. integrated for $\lambda > 912$\,\AA) is 
$L_{\rm n} = 1.1\times 10^{38}(d/1.6\kpc)^2$\es. 
%
%

\subsection{The first UV fluxes: Day 7.3 and beyond}

The first observation covering the UV was made at day 7.3 after 
the outburst. It showed that the maximum of the SED shifted to 
higher energies. The observed steep slope of the far-UV 
continuum with increasing fluxes for $\lambda <$1600\,\AA\ 
suggested $T_{\rm h} \gtrsim 50000$\,K. 
On the other hand, a flat horizontal profile of the continuum 
for $\lambda \gtrsim $2000\,\AA\ to the Balmer jump and its 
following gradual decrease to the IR domain suggested the 
presence of a strong nebular continuum produced by the thermal 
plasma. Therefore, subtracting the contribution from the giant, 
we can model the selected UV--near-IR fluxes with the function, 
$F(\lambda) = F_{\rm h}(\lambda) + F_{\rm n}(\lambda)$ 
(see Eq.~(11) of Paper~I). 

However, observing only the Rayleigh-Jeans tail of the hot star 
radiation, which dominates a very short wavelength region of 
$\sim$1150--1600\,\AA, did not allow to determine unambiguously 
its temperature from modelling the SED. 
To the contrary, the nebular component dominates a significant 
part of the available spectrum including its maximum at the 
near-UV, which allowed to determine its parameters ($T_{\rm e}$ 
and $EM$) unambiguously, within uncertainties of the observed 
fluxes (to $\lesssim 10$\,\%). 
Assuming that the nebular emission represents a fraction of 
the hot star (i.e. the burning WD) radiation, re-processed via 
the ionization/recombination events, we can determine only 
a lower limit of its temperature, $T_{\rm h}^{\rm min}$, at 
which the flux of ionizing photons just balances the rate of 
recombinations giving rise to the observed $EM$ (see Apendix~A). 

Accordingly, in fitting the UV/near-IR continuum at day 7.3 
to 19.5, I fixed the temperature of the ionizing source to 
$T_{\rm h}^{\rm min}$, which allowed to determine only 
limiting values of other depending parameters. 
For example, at day 7.3 the model SED gives 
$k_{\rm n} = 7.8\times 10^{17}$\,cm$^{-5}$ 
(i.e. $EM = 2.4\times 10^{62}(d/1.6\kpc)^2$\cmt) and 
$T_{\rm e} = 17000$\,K (i.e. $\alpha_{\rm B}({\rm H},T_{\rm e}) = 
1.67 \times 10^{-13}$ cm$^{3}$\,s$^{-1}$), 
which corresponds to $T_{\rm h}^{\rm min} = 110000$\,K for 
$F_{\rm h}$(1430\,\AA) = $3.6\times 10^{-10}$\ecsa\ 
(see Eq.~(\ref{eq:tmin2})). Consequently, the luminosity of 
the burning WD, 
$L_{\rm h} > L(T_{\rm h}^{\rm min}) = 2.2\times 10^{39}$\es, 
i.e. $> 13\,L_{\rm Edd}$ (see Eqs.~(\ref{eq:lth}) and 
(\ref{eq:lhgt}), Fig~\ref{fig:app}) and the effective radius 
of the ionizing source, $R_{\rm h}^{\rm eff} < 2.1$\ro. 

The \textsl{IUE} observation at day 19.5 was the last one prior 
to the emergence of a supersoft X-ray component 
\citep[see Fig.~1 of][]{ness+09}. At this day, the nebular 
radiation was still very strong having the 
$EM = 8.5\times 10^{61}(d/1.6\kpc)^2$\cmt\ and 
$T_{\rm e} = 19500$\,K. The far-UV continuum was estimated 
to $1.5\times 10^{-10}$\ecsa\ at $\lambda$ = 1280\,\AA, 
which yields 
  $T_{\rm h}^{\rm min} = 112000$\,K, 
  $R_{\rm h}^{\rm eff} < 1.1$\ro\ 
and 
  $L_{\rm h} > 6.9\times 10^{38}$\es, i.e. $> 4\,L_{\rm Edd}$. 

\subsection{The first supersoft X-rays: Day 26}

At day 26, it was the first possibility to model the SED of 
RS~Oph with the supersoft X-ray emission (see Sect.~2.5). 
Fitting the significant part of the total spectrum, from the 
supersoft X-rays (here for $\lambda > 20$\,\AA) to the IR 
domain, allowed to determine all 5 variables of the model 
SED unambiguously 
($\theta_{\rm h}$, $T_{\rm h}$, $N_{\rm H}$, $k_{\rm n}$ 
and $T_{\rm e}$; see Eq.~(11) and Sect.~4.2 of Paper~I). 
The best model ($\chi^2_{\rm red} = 2.8$ for 19 d.o.f.) 
corresponds to parameters of the stellar source 
$F_{\rm h}(\lambda)$: 
  $N_{\rm H} = (7.4\pm 0.4)\times 10^{21}$\cmd, 
  $T_{\rm h} = (310000\pm 10000)$\,K 
and 
  $\theta_{\rm h} = (8.4\pm 0.6)\times 10^{-12}$
 (i.e. $R_{\rm h}^{\rm eff} = (0.6\pm 0.04)(d/1.6\kpc)$\ro), 
which yields 
  $L_{\rm h} = (1.1\pm 0.22)\times 10^{40}(d/1.6\kpc)^2$\es. 
The nebular component, $F_{\rm n}(\lambda)$, is determined 
with 
  $k_{\rm n} = (2.0\pm 0.2)\times 10^{17}$\,cm$^{-5}$ 
  (i.e. $EM = (6.2\pm 0.6)\times 10^{61}(d/1.6\kpc)^2$\cmt) 
and 
  $T_{\rm e} = (20000\pm 3000)$\,K. 
The WD's pseudophotosphere thus dominated both the supersoft 
X-ray and the far-UV, whereas the nebular component dominated 
the remainder of the spectrum for $\lambda \gtrsim 2000$\,\AA\ 
(Fig.~\ref{fig:seds}). 
%
%
\begin{table*} 
\begin{center}
\scriptsize 
\caption{Physical parameters of the outbursting WD in RS~Oph 
         from day $\sim 1$ to day 26 
         (Sect.~3, Fig.~\ref{fig:seds}). 
        }
\begin{tabular}{cccccccccc}
\hline
\hline
~Day~                      &
~$N_{\rm H}$~              &
~$T_{\rm h}$~              &
~$\theta_{\rm h}$~         &
~$R_{\rm h}^{\rm eff}$~    &
~~~$\log(L_{\rm h}$)~~~    &
~$T_{\rm e}$~              &
~$EM$~                     &
~~$L_{\rm n}$~~            &
~~$\chi^2_{\rm red}$ / d.o.f. \\
~~                     &
~[$10^{21}$\cmd]~      &
~[K]~                  &
~~                     &
~[$R_{\odot}$]~        &
~[\es]~                &
~[K]~                  &
~[cm$^{-3}$]~          &
~[\es]~                &
                     \\
%
%
\hline
0.66--1.4& -- & $9600\pm 300$ 
        & $(2.2\pm 0.2)\times 10^{-9}$ & $156\pm 14$
        & $38.86\pm 0.11$ & ~~22000$^{\dagger}$ 
        & ~3.2$\times 10^{62}$ & ~$2.6\times 10^{38}$ & 2.9 / 10\\
3.8     & -- & $6700\pm 200$
        & $(2.3\pm 0.2)\times 10^{-9}$ & $163\pm 14$
        & $38.28\pm 0.10$ &~~22000 
        & ~1.3$\times 10^{62}$ & ~$1.1\times 10^{38}$ & 2.8 / 31\\
7.3     & -- & $> 110000^{\ddagger}$
        & $<2.9\times 10^{-11}$ & $< 2.1$ 
        & $>39.34$ &~~17000 
        & ~2.4$\times 10^{62}$ & ~$2.1\times 10^{38}$ & 1.6 / 11\\
14.6    & -- & $> 93000^{\ddagger}$
        & $<2.6\times 10^{-11}$ & $< 1.9$
        & $>38.96$ &~~16000 
        & ~9.8$\times 10^{61}$ & ~$8.7\times 10^{37}$ & 2.6 / 12\\
19.5    & -- & $>112000^{\ddagger}$
        & $<1.6\times 10^{-11}$ & $< 1.1$
        & $>38.84$ &~~19500 
        & ~8.5$\times 10^{61}$ & ~$7.2\times 10^{37}$ & 1.4 / 11\\
26.1    & 7.4$\pm 0.3$ & $310000\pm 10000$
        & $(8.4\pm 0.6)\times 10^{-12}$ & 0.60$\pm 0.04$
        & $40.05\pm 0.08$ &~~20000 
        & ~6.2$\times 10^{61}$ & ~$5.2\times 10^{37}$ & 2.8 / 20\\
%
%
\hline
\end{tabular}
\end{center}
$^{\dagger}$~fixed value, 
$^{\ddagger}$~= $T_{\rm h}^{\rm min}$ (Sect.~3.4, Appendix~A). 
\normalsize
\end{table*}
%

The spectral range between $\sim 6$ and $\sim 19$\,\AA\ was 
dominated by a flat harder component. 
This part of the X-ray spectrum was compared with a function 
\begin{equation}
  F_{\rm E} = k_{\rm E} E^{-\alpha} e^{-E/\beta} 
              e^{-\sigma_{\rm x} N_{\rm H}},
\label{eq:fE}
\end{equation}
where the flux of photons $F_{\rm E}$ at the energy $E$ (keV) 
is in 'photons/s/cm$^{2}$/keV', $k_{\rm E}$ is a scaling factor 
and $\alpha$, $\beta$ are fitting parameters. 
Figure~\ref{fig:seds} shows a model for 
$k_{\rm E} = 0.15$\,photons/s/cm$^{2}$/keV, 
$\alpha = 2.8$, 
$\beta = 5.5$\,keV and 
$N_{\rm H} = 7.4\times 10^{21}$\cmd. 
The absorbed flux of this harder (6.5--19.3\,\AA) component was 
$9.6\times 10^{-11}$\ecs, which corresponds to the unabsorbed 
value of $3.8\times 10^{-10}$\ecs. I note that the total 
(6.5--31.0\,\AA) absorbed flux of $1.3\times 10^{-10}$\ecs\ 
is comparable with that obtained by \cite{nelson+08}. 
The origin of this component is probably connected with the 
optically thin, a few millions of kelvins hot, 
collisionally-ionized plasma \citep[][]{bode+06}, 
heated by shock waves in the highly-supersonic mass-outflow. 

\section{Discussion}

\subsection{Biconical ionization structure during the first 4 days}

\begin{figure*}[!t]
\centering
\begin{center}
\resizebox{11cm}{!}{\includegraphics[angle=-90]{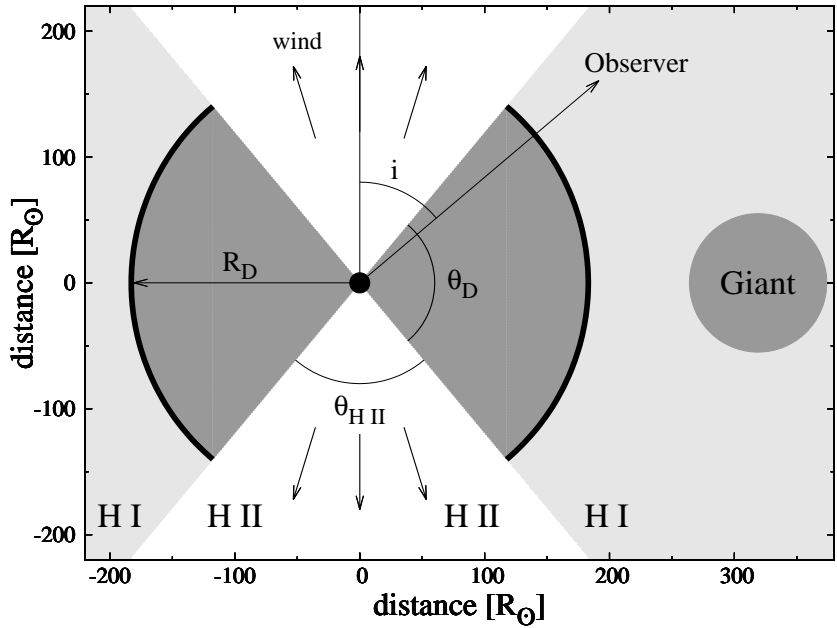}}
\end{center}
\caption[]{
A sketch of the ionization structure of RS~Oph during the first 
4 days after its explosion as seen on a cut perpendicular to 
the orbital plane containing the burning WD (the black circle). 
The neutral (H\,\I) zone is optically thick to the distance 
$R_{\rm D}$ from the WD, which represents the radius of the 
warm pseudophotosphere (heavy solid line). The ionized (H\,\I\I) 
zone has the opening angle $\theta_{\rm HII}$. Dimensions in 
the figure are given by the model SED and the elements of the 
spectroscopic orbit according to \cite{brandi+09}. 
          }
\label{fig:sketch}
\end{figure*}

Multiwavelength modelling of the SED revealed the simultaneous 
presence of a warm stellar ($F_{\rm h}(\lambda)$) and nebular 
($F_{\rm n}(\lambda)$) components of radiation in the spectrum 
of RS~Oph during the first four days after its explosion. 
The stellar component is characterized with 
$T_{\rm h} = 9600-6700$\,K and 
$L_{\rm h} = 7.2 - 1.9 \times 10^{38}$\es, 
which corresponds to production of 
$4.4\times 10^{45} - 1.1\times 10^{42}$ photons per second 
that are capable of ionizing hydrogen. 
However, the nebular component is characterized with 
$\textsl{EM} \sim 2.3 \times 10^{62}$\cmt, which corresponds 
to $\sim 3 \times 10^{49}$ recombinations per second 
$(\alpha_{\rm B}({\rm H},22000\,K) = 1.33\times 10^{-13})$, 
and thus cannot be generated by the observed stellar source 
of radiation 
($L_{\rm ph} \lll \alpha_{\rm B}\textsl{EM}$; see Eq.~(\ref{eq:lph})). 
This implies the presence of a hot ionizing source in the system, 
which is not seen directly by the outer observer. 
Such properties of radiative components were revealed in the 
spectrum of all symbiotic stars with a high orbital inclination 
during their outbursts. The corresponding spectrum was called 
a two-temperature-type \citep[see Sect.~5.3.4 of][]{sk05}. 
This puts constraints for the ionization and geometrical structure 
of the circumstellar matter surrounding the burning WD. 
According to \cite{sk05} the source of the warm stellar component 
can be ascribed to the WD's pseudophotosphere that is represented 
by the outer flared rim of an optically thick disk-like formation 
around the hot star. The circumstellar material located within 
the remainder of the sphere is ionized, and thus converts 
a fraction of the hot star (here the burning WD) radiation into 
the strong nebular emission (see Fig.~\ref{fig:sketch}). 

\subsubsection{Nature of the ionized region}

Ever since the very beginning of the nova explosion (day 1.16), 
very broad emission lines of hydrogen were observed in the 
optical spectrum \citep[e.g.][]{banerjee}. 
\cite{sk+08b} fitted the broad wings of \ha\ from day 1.38 
by the wind model of \cite{sk06}, in which a fraction of the wind 
from the central star is blocked by a disk around the WD's 
equator, and thus causes its bipolarity. 
The very high level of the continuum around the \ha\ line 
of $2.5\times 10^{-10}$\ecsa\ (see Fig.~\ref{fig:seds}, left 
top panel) yields its luminosity, $L_{\alpha}\sim 4900$\lo\ 
and the mass loss rate of a few 
$\times\,10^{-4}$\myr\ \citep[][]{sk+08b}. 
Consequently, the \ha\ luminosity was generated in a volume 
with the emission measure 
  \textsl{EM}$_{\alpha} = L_{\alpha}/\varepsilon_{\alpha} 
  \sim 10^{62}(d/1.6\kpc)^2$\cmt\ 
for $\varepsilon_{\alpha}(20000\,K) = 1.83\times 
10^{-25}$\,erg\,cm$^{3}$\,s$^{-1}$, which is consistent with 
that of the nebular continuum. 
\footnote{Its somewhat smaller value 
can be caused by a larger optical depth in the line and by 
contributions from singly and doubly ionized helium to the 
nebular continuum, which were not subtracted.} 
As both the nebular continuum and the hydrogen emission lines 
originate in the same region, the H\I\I\ zone, 
producing the strong nebular component of radiation in the SED, 
can be ascribed to the fast ionized wind from the outbursting WD. 

In addition, the biconical structure and the nature of the ionized 
zone was also supported by the presence of satellite emission 
components located at $\pm (2400-2500)$\kms\ with respect to the 
main cores of hydrogen lines, measured from the very beginning 
of the outburst and being recognizable 
to day $\sim 30$ \citep[][]{sk+08b,banerjee}. These emission 
features can be interpreted as a result of highly collimated 
bipolar mass outflow. 
This was confirmed by the radio 43\,GHz image from day 55 that 
showed directly highly collimated outflows produced by free-free 
emission from the thermal plasma \citep[][]{sok+08}. 
The presence of bipolar jets requires the presence of the 
inner disk, throughout which sufficient amount of mass must 
be accreted to balance the outflow via the jets 
\citep[e.g.][]{livio+03}. 
Thus the presence of a large disk-like formation encompassing 
the burning WD could also be responsible for keeping the 
super-Eddington luminosity of the burning WD for a long time. 
%
%
%

\subsubsection{Consequences of the bipolar structure}

(i) 
The bolometric luminosity of the burning WD cannot be determined 
unambiguously during this period of the nova evolution. Only its 
lower limit can be estimated from the components of radiation 
determining the observed two-temperature-type of the spectrum. 
The disk-like shell around the WD converts a fraction of its 
radiation into the warm stellar component (here approximated 
by blackbody). Its luminosity at day $\sim 1$ was 
$7.3\times 10^{38}$\es\ (Table~2). 
Another fraction of the burning WD's luminosity is reprocessed 
by ionization/recombination events into the nebular component. 
Its luminosity can be expressed as (see Eq.~(8) in Paper~I) 
\begin{equation}
 L_{\rm n} =  \textsl{EM} \int_{\lambda > 912}^{\infty} 
                          \varepsilon_{\lambda}(T_{\rm e}) 
                          {\rm d}\lambda ,
\label{eq:ln}
\end{equation}
where the nebula is assumed to radiate under conditions of Case~B. 
The \textsl{EM}(day $\sim 1) = 3.2\times 10^{62}$\cmt\ 
corresponds to $L_{\rm n} = 2.6\times 10^{38}$\es. This also 
represents a lower limit, because 
a fraction of ionizing photons is not detectable (they are not 
converted into the nebular photons, i.e. 
$L_{\rm ph} > \alpha_{\rm B}\textsl{EM}$, see Appendix~A) 
and, in addition, the nebula can be partially optically thick 
(i.e. the measured \textsl{EM} is smaller than that originally 
created by ionizations). Thus the bolometric luminosity of 
the burning WD is 
\begin{equation}
 L_{\rm h}^{\rm bol} > L_{\rm h} + L_{\rm n} = 
                       9.9\times 10^{38}\,{\rm erg}\,{\rm s}^{-1} 
\label{eq:lbol}
\end{equation}
at day $\sim 1$. Note that such adding of luminosities from 
different sources to the bolometric luminosity is possible only 
in the case of the two-temperature-type of the spectrum 
(i.e. the warm pseudophotosphere is far below the capability 
of giving rise to the observed nebular component of radiation). 

(ii)
According to the bipolar ionization structure of the burning WD, 
the vertical extension of the warm pseudophotosphere is limited 
by the orbital inclination to get hidden the ionizing source in 
direction of the observer. Its luminosity can be written as 
\begin{equation}
  S(R_{\rm D},\theta_{\rm D}) \sigma T_{\rm eff}^4 = 
  4\pi (R_{\rm h}^{\rm eff})^2\sigma T_{\rm eff}^4 ,
\label{eq:wlum}
\end{equation}
where the surface $S(R_{\rm D},\theta_{\rm D})$ of the 
pseudophotosphere is represented by a belt on a sphere with 
the radius $R_{\rm D}$, located around its equator and extended 
to the latitude $\pm \theta_{\rm D}/2$ (see Fig.~\ref{fig:sketch}). 
$R_{\rm h}^{\rm eff}$ is its effective radius 
(see Eq.~(6) in Paper~I). Then the radial extension of such the 
pseudophotosphere can be expressed as 
\begin{equation}
  R_{\rm D} = \frac{R_{\rm h}^{\rm eff}}
                   {\sqrt{\sin(\theta_{\rm D}/2)}}. 
\label{eq:rd}
\end{equation}
For $R_{\rm h}^{\rm eff} \sim 160$\ro\ (Table~2) and the 
limiting case $\theta_{\rm D}/2 = 90^{\circ} - i$, where 
$i = 50^{\circ}$, we get $R_{\rm D} = 200$\ro. 
Other possibilities, $40^{\circ} < \theta_{\rm D}/2 < 90^{\circ}$
yield $200 > R_{\rm D} > 160$\ro. 

\subsubsection{Comparison with other measurements}

At day 3.8 after the nova explosion, \cite{barry08b} used the KIN 
to find the structure of the circumstellar material around 
the star. They determined the angular size of the mid-infrared 
continuum emitting material to 4.0--6.2\,mas, which corresponds 
to linear size of 6.4--9.9(d/1.6\kpc)\,AU. 
As the N-band region is dominated by the nebular emission 
(see Fig.~\ref{fig:seds}), which is due to the fast ionized wind 
from the WD (see Sect.~4.1.1), its maximum extension (13\,AU for 
the velocity of 3000\kms\ and 3.8 days) is consistent with that 
detected by the KIN. 
The measurements obtained by the KIN in the inner spatial regime 
showed evidence of enhanced neutral atomic hydrogen emission and 
atomic metals located near the WD. This could reflect the H\,\I\ 
zone in the ionization structure inferred from the model SED 
(Fig.~\ref{fig:sketch}). 

\cite{monnier+06} measured the near-IR size of RS~Oph using 
the IOTA and Keck interferometer at the same day. They found 
a significantly asymmetric emission with the characteristic 
size of FWHM$\sim 3$\,mas. Observations for the following 
60 days did not indicate any change, suggesting thus that the IR 
emission is due to the $b-f$ and $f-f$ transitions in thermal 
plasma rather than from a hot gas in the expanding shock as 
considered in current popular models. Also these results -- the 
size and the nature of the IR emission -- are consistent with 
the model SED here. 

Using the AMBER/VLT interferometer, \cite{chesneau+07} measured 
the extension of the milliarcsecond-scale emission in the K band 
continuum, in the Br${\gamma}$ and He\,\I\,2.06\,$\mu$m lines 
at day 5.5. Observations showed that the emission from all 
investigated spectral regions is highly flattened with the ratio 
$b/a\sim 0.6$, having the same position angle and a typical 
Gaussian extension of $3.1\times 1.9$\,mas for the continuum. 

Thus, it is possible to conclude that the biconical ionization 
structure of the circumstellar matter around the WD in RS~Oph, 
as inferred from multiwavelength modelling the SED, is supported 
by direct interferometric measurements using IOTA, Keck and 
AMBER/VLTI performed during a few days after the eruption. 

\subsection{Sudden transition of the SED maximum to the UV}

The model SED for day 7.3 showed that the shift of the SED 
maximum from the optical to the far-UV (or even beyond) 
happened during a relatively short time of $< 3.5$ days 
(see Fig.~\ref{fig:seds}). A sudden decrease of the optical 
depth causing the shrinkage of the inflated WD from 
$\sim 160$\ro\ to less than 2.1\ro\ (Table~2) would require 
removal of a significant amount of dense, optically thick, 
material in the line of sight. The corresponding time of 
such the change should be proportional to the dynamical time 
scale of the inflated WD, 
\begin{equation}
          t_{\rm dyn} = (G \rho)^{-1/2} ,
\label{eq:tdyn}
\end{equation}
where $\rho$ is the density of the matter giving rise to the 
gravitational force. So, for the WD mass of $1.3$\mo, its 
effective radius 160\ro\ would collapse during 
$t_{\rm dyn} = 67$ days, which is by a factor of $\sim 20$ 
larger than the time scale of the observed change in the 
effective radius. Therefore, a gradual shrinking of the 
WD pseudophotosphere due to the expansion of the outflowing 
material cannot be the cause of its sudden disappearance 
between day 3.8 and 7.3. 

The sudden change in the SED can be a result of a change 
in the position of the ionization boundary with respect 
to the line of sight. If the opening angle of the H\,\I\ 
zone $\theta_{\rm D}$ (see Fig.~\ref{fig:sketch}) shrinks 
so that $\theta_{\rm D}/2 < 90^{\circ} - i$, the burning WD 
will rise up the horizon represented by the H\,\I/H\,\I\I\ 
boundary. A shrinkage/enlargement of the 
$\theta_{\rm D}/\theta_{\rm HII}$ angle can be a result of 
a decreasing mass loss rate from the WD. Recently, \cite{cask12} 
found that the ionization structure of the hot components in 
symbiotic binaries during outbursts is similar to that 
sketched in the figure~\ref{fig:sketch} here (see their Figs.~1 
and 6). In their model, the opening angle of the H\,\I\I\ zone, 
$\theta_{\rm HII}$, is related to the mass loss rate from the 
WD as $\dot M_{\rm wind}^{-2}$. This implies that a relatively 
small decrease in the $\dot M_{\rm wind}$ can yield a large 
increase in the $\theta_{\rm HII}$ angle 
\citep[see Fig.~1 of][]{cask12}. 
During the first days, when the $\dot M_{\rm wind}$ was at 
a maximum, $\theta_{\rm HII}/2 < i$ (i.e. the hot WD was below 
the H\,\I/H\,\I\I\ boundary), because we observed the warm 
pseudophotosphere. 
A small decrease in $\dot M_{\rm wind}$ could lead to opening 
of the H\,\I\I\ zone so that $\theta_{\rm HII}/2 > i$, i.e. 
the burning WD raised up the H\,\I/H\,\I\I\ horizon and thus we 
could observe directly its hot $\lesssim 2.1$\ro\ photosphere 
at day 7.3. This could happen in a short time, because of 
a relatively low value of $i$. A transient increase of 
the \textsl{EM} at day 7.3 supports this view. 

\subsection{Long-lasting super-Eddington luminosity}

Multiwavelength modelling the SED revealed that the luminosity 
of the WD was super-Eddington during the whole period of its 
nuclear burning (Table~2 here and Table~2 of the following 
paper). 

Prior to the emergence of supersoft X-rays, only the lower limit 
of the WD's luminosity could be determined. During the presence 
of the two-temperature-type of the spectrum (i.e. for the first 
4 days), it was given by the integral of directly measured 
fluxes (or their model) scaled with the distance. The model SED 
at day $\sim 1$ and 3.8 corresponded to the luminosity of 6 and 
2\,$L_{\rm Edd}$, respectively, for $M_{\rm WD} = 1.3$\mo. 
However, its true value was higher, because both the warm shell 
and the nebula converted only a fraction of the WD radiation 
into direction of the observer (Sect.~4.1.2, Eq.~(\ref{eq:lbol}), 
Fig.~\ref{fig:sketch}). 

From day 7.3, measuring only the short Rayleigh-Jeans tail of 
the hot WD radiation, only the lower limit of its temperature 
and luminosity could be determined (see Appendix~A). 
Using the emission measure from the model SED and the 
$F_{\rm h}(\lambda)$ flux from the far-UV spectrum, allowed 
to determine the minimum temperature of the WD photosphere, 
$T_{\rm h}^{\rm min} \sim 10^5$\,K (Eq.~(\ref{eq:tmin2}), 
Table~2) and the minimum of its luminosity, 
$L(T_{\rm h}^{\rm min}) = 13 - 4\,L_{\rm Edd}$ 
(Eq.~(\ref{eq:lth})), at which the burning WD is just capable 
of producing the observed \textsl{EM} during day 7.3 -- 19.5. 
Note that the luminosity of the nebular component itself 
corresponds to 1.24\,$L_{\rm Edd}$ at day 7.3 (Table~2). 

The emergence of the first supersoft X-rays from the WD at 
day 26 allowed to determine its luminosity unambiguously 
to $L_{\rm h} = 65\pm 12$\,$L_{\rm Edd}$ (Sect.~3.5). 
Such a highly super-Eddington luminosity contradicts theoretical 
predictions, treating the nova event as a simple nuclear 
bomb explosion. 
According to \cite{starrfield+08} the luminosity can exceed 
the Eddington limit only at the very beginning of the nova 
eruption with a rapid decline below it on a time-scale of hours. 

However, the opposite cases were reported by some authors 
already during 70's and 80's decades of the last century. 
For example, \cite{friedjung87} showed that the energy output 
of the nova FH~Ser was well above the Eddington limit for 
about two months after its optical maximum. The super-Eddington 
state was theoretically investigated by \cite{shaviv01}, who 
considerred a reduction of the effective opacity due to 
the rise of a `porous layer' above the convective zone of 
the burning WD \citep[see Fig.~1 of][]{shav+dot10}. These 
conditions enlarges the Eddington luminosity well above 
its classical value, calculated for the Thomson scattering 
opacity. The super-Eddington state also accelerates a thick 
continuum driven wind at high rates. 
The mass loss through the wind at a high rate of a few 
$\times\,10^{-4} - 10^{-5}$\myr\ during the hydrogen burning 
period of RS~Oph \citep[][]{sk+08b} is consistent with 
that suggested for the super-Eddington state 
\cite[Eq.~(30) of][for $v_{\rm s}$ = 20 - 50\kms]{shaviv01}. 
%
In addition, the presence of the bipolar jets for the first 
30 days \citep[][]{sk+08b,banerjee}, which suggests accretion 
throughout the disk at a high rate, also supports 
the super-Eddington state. 
For example, accretion at $\dot m \sim 2\times 10^{-5}$\myr\ 
can release its binding energy 
$G M_{\rm WD} \dot m/R_{\rm WD} \sim 10^{39}$\es, and thus 
help to sustain the high luminosity of the accretor for a 
long time. 

Finally, it is important to note that the high energy output 
released during the outburst of RS~Oph, as given by the global 
model SED of this paper, is more than a factor of 10 larger 
than that so far derived. This requires a very high accretion 
rate, which is far above that can be obtained from a spherical 
wind mass loss of the giant \citep[e.g.][]{schaefer09}. 
Considering that the primary source of the material powering 
the outburst is the wind from the giant, then a different, more 
efficient mechanism of the mass transfer onto the WD must be 
in the effect. 
This could be realized by the so-called wind Roche-lobe overflow 
(WRLOF), which can occur when the wind acceleration zone is 
stretched out to the Roche lobe of the giant 
\citep[e.g.][]{moh+pod07,abate+13}. 
In this case the wind of the giant is focused towards the 
orbital plane and in particular towards the WD. The accretion 
rate predicted in the WRLOF regime can be 100 times larger 
than the rate expected from standard Bondi-Hoyle accretion 
\citep[see][]{moh+pod07}. 

More detailed discussion on the accretion mechanism is presented 
in the following paper, {\em RS~Ophiuchi: The supersoft X-ray 
phase and beyond}, where other critical parameters are determined 
from the model SEDs during the SSS phase and following quiescence. 

\section{Summary}

In this paper I applied the method of multiwavelength modelling 
the SED (see Paper~I) to the recurrent symbiotic nova RS~Oph 
from its explosion to the SSS phase (to day 26). The model SEDs 
revealed the presence of a strong stellar as well as nebular 
component of radiation in the spectrum. The contribution from the 
cool giant was negligible with respect to these components 
(Sect.~3, Fig.~\ref{fig:seds}, Table~2). The main results can 
be summarized as follows. 
\begin{enumerate}
\item
The radiation from the giant corresponds to its effective temperature 
  $T_{\rm eff} = 3800 - 4000$\,K, 
the radius, 
  $R_{\rm g} = (61 - 55)\,(d/1.6\kpc)$\ro\ 
and the luminosity, 
  $L_{\rm g} = (690 \pm 95)\,(d/1.6\kpc)^2$\lo. 
The giant underfills deeply its Roche lobe for d = 1.6\kpc\ 
(Sect.~3.1). 
\item
During the first 4 days after the explosion, the model SED 
identified a biconical ionization and geometrical structure
of the nova. 
The $\sim 8200$\,K warm pseudophotosphere was represented by 
a belt on a sphere with the radius of 160--200\ro, located around 
the WD's equator and extended to the latitude $> 40^{\circ}$. 
The remaining space around the WD's poles was ionized, producing 
a strong nebular emission 
($\textsl{EM} \sim 2.3 \times 10^{62}$\cmt) via the fast ionized 
wind from the WD (see Fig.~\ref{fig:sketch}, Sect.~4.1). 
\item
The luminosity of the burning WD was super-Eddington for the 
whole investigated period, although it was possible to determine 
only its lower limit prior to the emergence of the supersoft 
X-ray radiation (Sect.~4.3, Table~2). 

During the first 4 days, the luminosity 
$L_{\rm h} > 2-6\,L_{\rm Edd}$ for $M_{\rm WD} = 1.3$\mo. 
Here the lower limit is given by the fit of the model SED to 
the total measured spectrum, because both the neutral and the 
ionized region convert only a fraction of the WD's radiation 
into direction of the observer (Sect.~4.1.2). 

From day 7.3, the two-temperature-type of the spectrum suddenly 
disappeared, because the H\,\I\I\ zone opened so that we could 
observe directly the hot WD (Sect.~4.2). In this case lower 
limits of the WD's temperature and luminosity were determined 
from the condition that the corresponding flux of ionizing 
photons was just capable of producing the observed \textsl{EM} 
(Sect.~4.3, Eq.~(\ref{eq:lph})). Here the model SEDs suggested 
$L_{\rm h} > 4-13\,L_{\rm Edd}$. 

At day 26, the first detection of the supersoft X-ray fluxes 
allowed to determine the WD's luminosity unambiguously 
to $L_{\rm h} = 65\pm 12$\,$L_{\rm Edd}$ 
(Sect.~3.5, Fig.~\ref{fig:seds}). 
\item
The long-lasting super-Eddington luminosity and the mass loss 
at a very high rate are consistent with theoretical considerations 
treating the super-Eddington state as a result of a reduction 
the effective opacity of the WD atmosphere. 
However, the appearance of highly super-Eddington luminosity 
just at the beginning of the SSS phase (day 26) is enigmatic, 
unless this could also be connected with a relevant reduction 
of the effective opacity. 
The presence of collimated jets during the burning phase suggests 
an accretion throughout a disk during this period, which can help 
to balance the high energy output for a long time. This means 
that the disk was not destroyed by the eruption. 
%
\item 
The high energy output released during the outburst of RS~Oph 
requires a very high accretion rate, which could be realized in 
the WRLOF regime. 
\end{enumerate}
Further analysis and discussion on the accretion mechanism is 
found in the following paper, {\em RS~Ophiuchi: The supersoft 
X-ray phase and beyond}.

\section*{Acknowledgments}
\textsl{IUE} spectra presented in this paper were obtained 
from the MAST. 
The author thanks Manfred Hanke for extracting the 
\textsl{XMM-Newton} observation from the archive and 
Richard Barry for valuable discussion about the Keck 
Interferometer Nuller observing regimes. 
This research has been in part supported by the project     
No.~SLA/103115 of the Alexander von Humboldt foundation 
and by the Slovak Academy of Sciences under 
a grant VEGA No.~2/0002/13.

\appendix
 
\section{A minimum $T_{\rm h}$ and $L_{\rm h}$ of the ionizing 
         source from the observed \textsl{EM}}

\begin{figure}
\centering
\begin{center}
%
\resizebox{\hsize}{!}{\includegraphics[angle=-90]{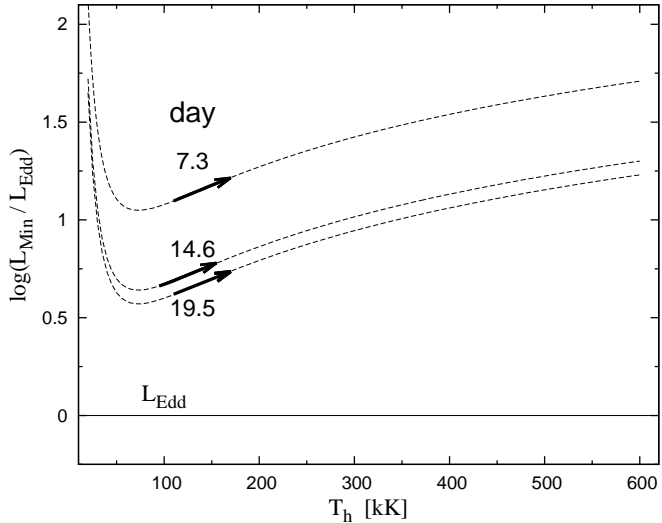}}
\end{center}
\caption[]{ 
The lower limit of the WD luminosity as a function of its 
temperature scaled with \textsl{EM} (Eq.~(\ref{eq:lth})) 
derived from the model SED on day 7.3, 14.6 and 19.5. The 
thick arrows start at $T_{\rm h}^{\rm min}$ (Table~2). 
          }
\label{fig:app}
\end{figure}

Here we determine the lower limit of the temperature and 
luminosity of the ionizing source, at which the corresponding 
flux of ionizing photons gives rise to the observed $EM$. 
The method is based on the fact that the nebular component 
of radiation represents a fraction of the hot stellar radiation 
reprocessed via the ionization/recombination events. 
The result of this process depends on the number of ionizing 
photons produced by the hot star and the number of particles 
on their path that are subject to ionization. 
In the limiting case, when {\em all} the ionizing photons are 
converted into the diffuse radiation, we can approximate the 
equilibrium condition between the flux of ionizing photons 
$L_{\rm ph}$ and the rate of recombinations in the nebula as 
%
%
\begin{equation}
 L_{\rm ph}({\rm H})\, =\, \alpha_{\rm B}({\rm H},T_{\rm e})\,
                           \textsl{EM}, 
\label{eq:lph}
\end{equation}
where $\alpha_{\rm B}({\rm H},T_{\rm e})$ (cm$^{3}$\,s$^{-1}$) 
is the recombination coefficient to all but the ground state 
of hydrogen (i.e. for Case $B$). 
This equation is valid for the hydrogen plasma, heated by
photoionizations, and characterized with a constant
$T_{\rm e}$ and particle concentration. Thus, the $EM$ 
constrains a minimum luminosity of $L_{\rm ph}$ photons,
which is a function of $T_{\rm h}$ and $L_{\rm h}$ of the hot 
ionizing source. Using the expression for the $L_{\rm ph}$ 
function \citep[see Eq.~(11) of][]{sk01}, the luminosity of 
the ionizing source, which is capable of producing the observed 
\textsl{EM}, can be written as 
%
%
\begin{equation}
 L(T_{\rm h}) = \alpha_{\rm B}({\rm H},T_{\rm e})\,\textsl{EM}
                \frac{\sigma T_{\rm h}^{4}}{f(T_{\rm h})},
\label{eq:lth}
\end{equation} 
%
where the function
%
%
\begin{equation}
f(T_{\rm h}) = \frac{\pi}{hc}\int^{\rm 912\AA}_{0}\!\!\!
               \lambda\, B_{\lambda}(T_{\rm h})\,\rm d\lambda.
\label{eq:fth}
\end{equation}
Examples of the $L(T_{\rm h})$ function parametrized with 
different values of \textsl{EM} are plotted in Fig.~\ref{fig:app}. 
Its minimum corresponds to the temperature, at which the source 
produces maximum flux of ionizing photons (for hydrogen it is 
73000\,K). For higher/lower $T_{\rm h}$ it increases, because 
of decreasing production of the ionizing photons for 
a given $L_{\rm h}$. 

For $T_{\rm h} \gtrsim 50000$\,K we can observe only the Jeans 
tail of the hot source radiation in the ultraviolet. This implies 
that a small part of the far-UV fluxes increasing to higher 
energies can be compared with any blackbody radiation 
from this range of temperatures. If we have available an 
appropriate spectrum, a $Zanstra$-tepmerature can be adopted. 
Here, according to Eq.~(\ref{eq:lph}), we determine a minimum 
temperature $T_{\rm h}^{\rm min}$ of the ionizing source, at 
which the blackbody radiation, scaled to the far-UV fluxes, is 
just capable of producing the observed \textsl{EM}. 
According to \cite{sk05} we can get $T_{\rm h}^{\rm min}$ by 
solving equation
%
%
\begin{equation}
  \frac{k_{\rm n}}{k_{\rm h}(T_{\rm h})} 
                  \alpha_{\rm B}({\rm H},T_{\rm e})
                   - f(T_{\rm h}) = 0 ,
\label{eq:tmin1}
\end{equation}
where $k_{\rm n}$ and $T_{\rm e}$ are fitting parameters (see 
Sect.~2.4 of Paper~I) and $k_{\rm h} (= \theta_{\rm h}^2)$ 
scales the blackbody flux emitted by the ionizing source to its 
observed flux $F_{\rm h}(\lambda)$. 
Then, Eq.~(\ref{eq:tmin1}) can be expressed as 
%
%
\begin{equation}
  \frac{k_{\rm n}}{F_{\rm h}(\lambda)}\alpha_{\rm B}
       ({\rm H},T_{\rm e})\pi B_{\lambda}(T_{\rm h}) 
       - f(T_{\rm h}) = 0 ,
\label{eq:tmin2}
\end{equation}
where $F_{\rm h}(\lambda)$ is selected at a wavelength $\lambda$, 
at which the ionizing source dominates the spectrum (usually in 
the far-UV). Its solution for the input quantities of $k_{\rm n}$, 
$F_{\rm h}(\lambda)$, $\lambda$ and 
$\alpha_{\rm B}({\rm H},T_{\rm e})$ provides 
$T_{\rm h} \equiv T_{\rm h}^{\rm min}$. 

According to Eq.~(\ref{eq:lth}), this temperature determines 
the minimum luminosity $L(T_{\rm h}^{\rm min})$, at which the 
ionizing source is just capable of producing the observed 
\textsl{EM}. 
In the real case, a fraction of the ionizing photons can escape
the star without being converted into the nebular radiation.
Also the measured \textsl{EM} represents a lower limit of what 
was originally created by ionizations, because we observe only 
the optically thin part of the nebula. Under these circumstances 
the luminosity of the $L_{\rm ph}$ photons, originally emitted 
by the source, is larger than that given by the equilibrium 
condition (\ref{eq:lph}), i.e. 
$L_{\rm ph} > \alpha_{\rm B}({\rm H})\,$\textsl{EM}, which
according to Eq.~(\ref{eq:lth}) implies 
\begin{equation}
   L_{\rm h} > L(T_{\rm h}^{\rm min})
\label{eq:lhgt}
\end{equation}
even for $T_{\rm h} = T_{\rm h}^{\rm min}$. Considering that 
$T_{\rm h}$ can be $> T_{\rm h}^{\rm min}$ and the nebula during 
outbursts is ionization bounded, it is probable that 
$L_{\rm h} \gg L(T_{\rm h}^{\rm min})$. 
%
%
\end{document}